\documentclass{accel}

\newcommand {\be}{\begin{equation}}
 \newcommand {\ee}{\end{equation}}
 \newcommand {\bea}{\begin{eqnarray}}
 
 \newcommand {\eea}{\end{eqnarray}}

 \renewcommand {\theequation}{\thesection.\arabic{equation}}

\usepackage[dvips]{graphicx}
\usepackage{latexsym}
\usepackage{amsbsy}
\usepackage{lineno}

\begin{document}

\title{Clear indication of a strong $I$=0 $\bar KN$ attraction in the $\Lambda (1405)$ region  from the CLAS photo-production data }

\author{
\Name{M. Hassanvand} \institute*{*1}
\Name{Y. Akaishi} \institute*{*2}
\Name{T. Yamazaki} \institute*{*2,*3}
}
\INSTITUTE{1}{Department of Physics, Isfahan University of Technology}
\INSTITUTE{2}{RIKEN, Nishina Center}
\INSTITUTE{3}{Department of Physics, University of Tokyo}

\maketitle

Possible existence of deeply bound kaonic nuclear systems was proposed \cite{Akaishi02} more than a decade ago, based on an ansatz that the $\Lambda^* \equiv \Lambda(1405)$ mass is 1405 MeV$/c^2$, where the $\Lambda^*$ is a $\bar{K} N$ quasi-bound state decaying to $\Sigma\pi$. 
Recently, a large number of data on photo-production of $\Lambda(1405)$ in the $\gamma p \to K^+ \pi^{0\pm} \Sigma^{0\mp}$ reaction were provided  by the CLAS collaboration \cite{Moriya13}, and  double-pole structure of the $\Lambda^*$ has been intensively discussed by chiral dynamics analyses \cite{Roca13,Mai15,Meissner15}. In contrast, we show a $\Lambda^*$ mass of 1405 MeV/$c^2$ is deduced from the same CLAS data.

\begin{figure}[htb] 
\begin{center}
\includegraphics[width=8cm]{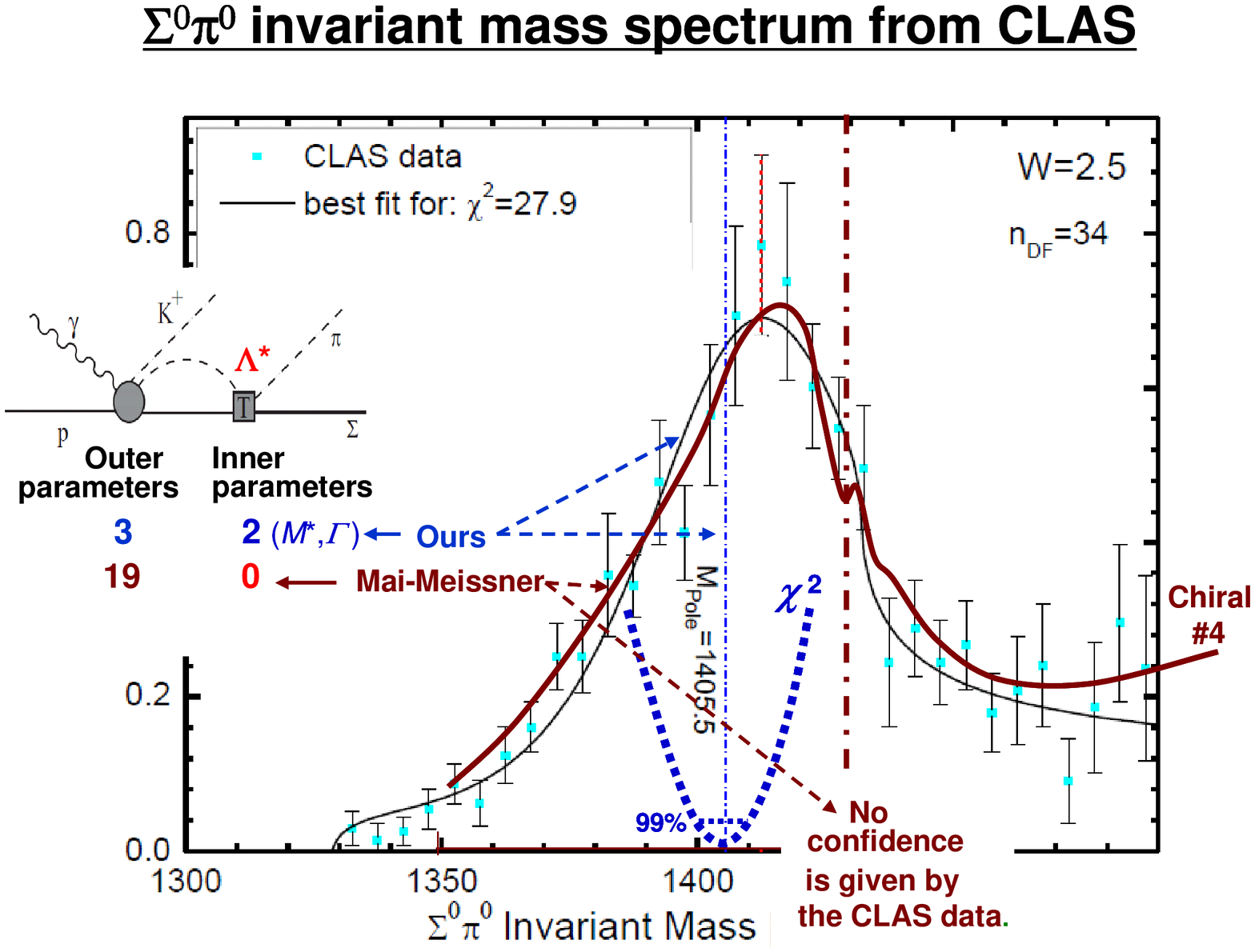}
\vspace{0.0cm}
\caption{Analyses of a $\Sigma^0 \pi^0$ invariant mass spectrum from CLAS. Mai-Meissner's analysis (brown) with chiral theory and Hassanvand-Akaishi-Yamazaki's analysis (black and blue) are compared. }
\label{CLAS}
\end{center}
\end{figure} 

We have analyzed the CLAS data. Figure \ref{CLAS} compares two kinds of  CLAS-data analyses. The most essential question is what the pole position of $\Lambda^*$ extracted from the CLAS data themselves is. To answer it, we classified $\chi^2$ fitting parameters into "inner" and "outer" ones, where the "inner" indicates the parameters appearing inside the $T$-matrix which can vary the pole position and width of $\Lambda^*$. The double pole positions recommended in ref. \cite{Meissner15} are cases selected by using only "outer" parameters, holding the pole position unchanged in $\chi^2$ fitting processes. Therefore, the chiral pole position in ref. \cite{Mai15}, for example, gets no confidence from the CLAS data. On the other hand, we have used the $\Lambda^*$ pole position and width as fitting parameters and found the $\chi^2$ minimum at 1405.5 MeV$/c^2$, rejecting the chiral result with more than 99\% statistical significance.

\begin{figure}[htb] 
\begin{center}
\includegraphics[width=7cm]{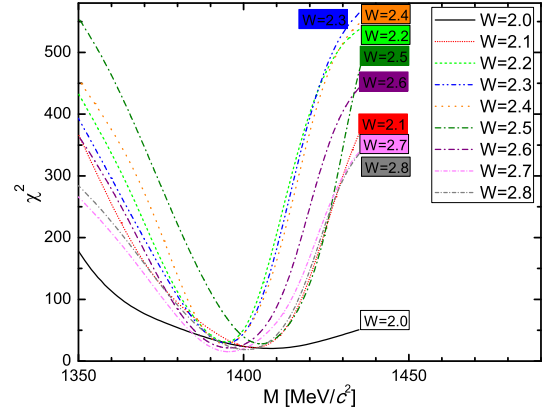}
\vspace{0.0cm}
\caption{Variation of $\chi^2$ values with respect to $\Lambda^*$ pole position parameter, $M$, for the $\Sigma^0 \pi^0$ invariant mass spectra at $\gamma p$ total energies, $W=2.0 \sim 2.8$ GeV, from CLAS. The $\chi^2$ minimum appears around 1400 MeV$/c^2$.}
\label{chi2}
\vspace{-0.5cm}
\end{center}
\end{figure} 

Figure \ref{chi2} shows the variation of $\chi^2$ values with respect to the $\Lambda^*$ pole position for all the $\Sigma^0 \pi^0$ invariant mass data from CLAS. The $\chi^2$ minimum, $\chi^2_{\rm min}(M=M_{\rm pole})$, appears around $M_{\rm pole}=1400$ MeV$/c^2$ in all cases. A statistical confidence of the $\Lambda^*$ pole position, $M_{\rm pole}$, can be obtained through the increment of $\Delta \chi^2(M) \equiv \chi^2(M)-\chi^2_{\rm min}(M_{\rm pole})$: $\Delta \chi^2=2.36, 4.74, 9.23$, corresponding to confidence levels of 68.3, 95 and 99.9\%, respectively. Thus, the confidence of the $\Lambda^*$ pole position, that is of our main concern, was obtained through the "inner" fitting parameters. It is again stressed that the "outer" parameters used in \cite{Mai15} (see also Table in \cite{Meissner15}) cannot get any {\it quantitative} confidence about the $\Lambda^*$ pole position, in spite of the beautiful reproduction of all the global neutral and charged spectra of CLAS.  

In summary, the pole position for $\Lambda(1405)$ extracted from the CLAS photo-production is 
not shallow-$\bar KN$-binding ones, $1421 \sim 1434$ MeV$/c^2$  ~\cite{Meissner15}, but is consistent with the deep-$\bar KN$-binding PDG value, $1405.1^{+1.3}_{-1.0}$ MeV$/c^2$.

\end{document}